\documentclass[runningheads]{llncs}
\usepackage[T1]{fontenc}
\usepackage{tikz}
\usetikzlibrary{arrows.meta, positioning}

\usepackage{caption}
\usepackage{graphicx}
\usepackage{amsmath}
\usepackage[textsize=small]{todonotes}
\usepackage{siunitx}
\usepackage{hyperref}

\begin{document}

\title{Compressing ACAS-Xu Lookup Tables with Binary Decision Diagrams}
%
%
\author{Martin Boniol\inst{1,2}\orcidID{0009-0005-0544-6778} \and
Julien Brunel\inst{2}\orcidID{0009-0004-3639-6681} \and
Jean-Baptiste Chaudron\inst{1}\orcidID{0000-0002-2142-1336} \and
Christophe Garion\inst{1}\orcidID{0000-0002-4467-2939} \and
Xavier Thirioux\inst{1}\orcidID{0009-0002-1126-6835}}
\authorrunning{F. Author et al.}
%
\institute{Université de Toulouse, ISAE SUPAERO, Toulouse, France \and
Université de Toulouse, DTIS, ONERA, Toulouse, France}

%
\maketitle              

\emph{This preprint has not undergone peer review or any post-submission improvements or corrections.
The Version of Record of this contribution is published in NFM 2026, and will be available online.}

\begin{abstract}

The Airborne Collision Avoidance System Xu (ACAS-Xu) relies on large certified Look-Up Tables (LUTs) that encode the exact decision logic used in operation.
Neural-network-based approximations have been proposed to reduce memory requirements, but they inherently introduce approximation errors and complicate formal verification.
This paper presents a symbolic compression approach based on Binary Decision Diagrams (BDDs) that preserves the exact semantics of the ACAS-Xu LUTs.
The resulting representation is canonical, deterministic, and fully equivalent to the original tables, enabling sound and exact reasoning over the complete decision logic.
By expressing both the system behavior and domain-specific operational properties within a common Boolean framework,
verification reduces to efficient BDD operations and emptiness checks, with precise counterexamples generated when properties are violated.
We demonstrate that the proposed BDD-based representation significantly reduces memory usage, achieves predictable and low-latency execution,
and can be deployed on embedded platforms.
These results highlight BDDs as a compelling alternative for exact, verifiable, and embedded deployment of ACAS-Xu decision logic.

\keywords{Binary Decision Diagram \and Formal Verification \and ACAS-Xu \and Neural Network \and Embedded System \and UAV.}
\end{abstract}
\section{Introduction}

Ensuring safe separation between aircraft is a critical challenge in increasingly dense airspaces.
The ACAS-X addresses this challenge through a family of collision avoidance logics, including ACAS-Xu,
a variant designed specifically for Unmanned Aerial Vehicles (UAVs) and supporting both vertical and horizontal maneuvers \cite{GuideAcas,Mops}.

The ACAS-Xu decision logic is implemented using large LUTs.
While LUTs guarantee deterministic behavior, their memory footprint poses a serious obstacle to deployment on UAV platforms.
In particular, the horizontal avoidance tables alone require several gigabytes of storage, challenging the capabilities of typical onboard avionics \cite{UAS}.
Moreover, accessing such large tables introduces non-trivial latency and complicating real-time execution on resource-constrained hardware.

To alleviate this issue, prior work has explored compressed
representations of the ACAS-Xu logic, most notably using Neural
Networks (NN)~\cite{DnnCompression}.
While such approaches significantly reduce memory requirements, they rely on approximation and therefore do not preserve the exact semantics of the original LUTs,
raising concerns for safety-critical deployment~\cite{Unsafe,SafetyNet}.

In this work, we propose a new compression technique for ACAS-Xu based on Binary Decision Diagrams (BDDs).
BDDs provide a canonical and exact symbolic representation of discrete decision functions,
enabling substantial reductions in memory footprint while preserving the precise behavior of the original LUTs~\cite{Bryant1986,Clarke1999}.
Beyond compactness, the symbolic nature of BDDs enables efficient reasoning about the decision logic,
facilitating the verification of safety and consistency properties that are difficult to analyze directly on large LUTs or NNs.
We demonstrate that the resulting BDD-based representation can be evaluated in real time on resource-constrained platforms,
including both CPU-based and FPGA-accelerated implementations.

The rest of this paper is organized as follows:
Section~\ref{sec:acasxu} introduces the ACAS-Xu system and Section~\ref{sec:NN} presents a compression strategy using NNs.
Section~\ref{sec:BDD} details the BDD-based compression of the ACAS-Xu LUTs.
In Section~\ref{sec:verif} we present a strategy to verify our BDDs.
Section~\ref{sec:expe} details the comparaison between NNs, BDDs and LUTs through experimental evaluation of each implementations on embedded systems.
Finally, Section~\ref{sec:conclusion} concludes and outlines perspectives for future work.


\section{ACAS-Xu presentation}
\label{sec:acasxu}

\subsection{Decision logic and advisories}

\begin{figure}[h!]
    \centering
    \includegraphics[width=6cm]{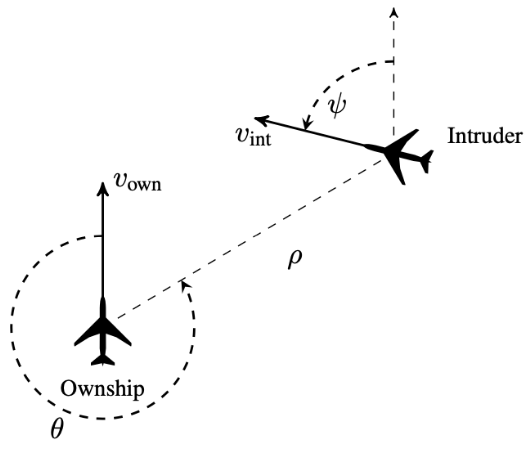}
    \caption{ACAS-Xu geometry (taken from~\cite{Reluplex})}
    \label{fig:geometry}
\end{figure}

ACAS-Xu is the collision avoidance system designed for unmanned aircraft within the ACAS-X family \cite{GuideAcas}.
It determines whether an avoidance maneuver is required by evaluating the relative geometry between the ownship and surrounding traffic.
The system operates on a seven-dimensional state vector derived from sensor measurements and relative kinematics, as illustrated in Figure~\ref{fig:geometry}.
This state is defined by the following parameters:
$\tau$ (time until loss of vertical separation), $\rho$ (distance between the ownship and the intruder),
$\theta$ (angle to the intruder relative to the ownship’s heading direction), $\psi$ (intruder's heading angle relative to the ownship's heading), 
$v_{own}$ (ownship speed), $v_{int}$ (intruder speed) and $a_{prev}$ (previous advisory issued by the system).

Based on this state representation, ACAS-Xu selects an appropriate avoidance resolution advisory.
For horizontal conflict resolution, the system can issue one of five possible advisories :
Clear of Conflict (\emph{COC}), Weak Left (\emph{WL}), Weak Right (\emph{WR}), Strong Left (\emph{SL}) and Strong Right (\emph{SR})

\subsection{LUTs as the Core Decision Representation}

The ACAS-Xu decision logic relies on several structural features that characterize how the system evaluates potential collision scenarios.
When an advisory must be issued, the system first acquires the current encounter state, which describes the relative geometry and kinematics of the ownship and the intruder.
This continuous state is then discretized according to the predefined ACAS-Xu quantization grids, and the corresponding LUT entry is selected by mapping the state to its nearest discretized point \cite{Mops,GuideAcas}.

\begin{figure}[h!]
    \centering
    \includegraphics[width=10cm]{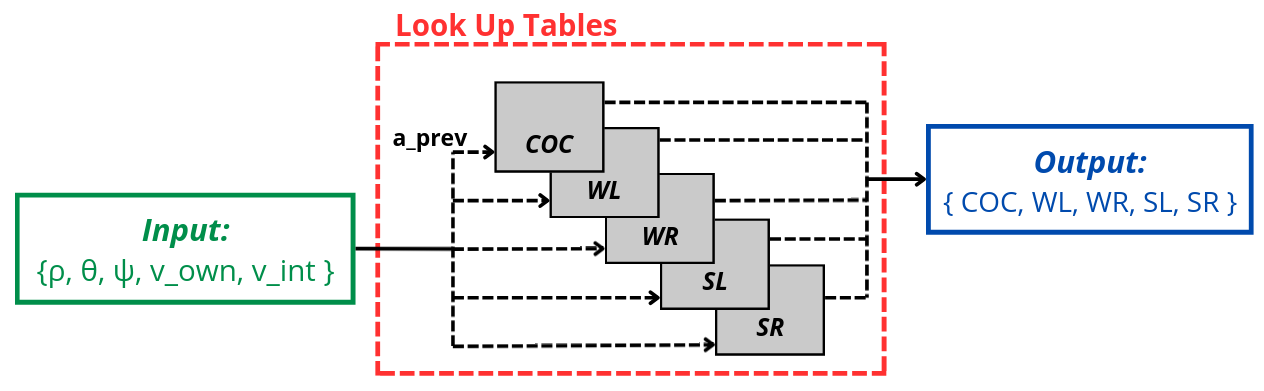}
    \caption{ACAS-Xu Architecture}
    \label{fig:LUT}
\end{figure}

For each discretized encounter state, the LUT stores the expected cost associated with every possible advisory, as computed offline using Markov Decision Processes (MDPs) \cite{Kochenderfer2011DP}.
These advisory costs are all returned by the LUTs but the advisory minimizing this expected cost is selected to be the advisory provided to the pilot.
Each LUT covers nearly 90 million discretized encounter states, representing a comprehensive set of possible relative configurations under varying geometries, speeds, and sensor uncertainties.
In addition, the system maintains five distinct LUTs, one for each value of the previously issued advisory ($a_{prev}$). The global architecture is shown in Figure~\ref{fig:LUT}.

\section{Neural Network–based compression and its limitations}
\label{sec:NN}

\subsection{The Reluplex NNs}

Neural networks have been widely investigated as a way to compress the ACAS-Xu LUTs,
primarily due to their ability to approximate high-dimensional decision functions with a substantially smaller memory footprint~\cite{Julian2019PolicyCompression}.

We report here experimental results obtained with the ACAS-Xu networks originally introduced by Julian et al~\cite{DnnCompression}.
These models constitute the most widely used reference implementation in the literature,
and evaluating them offers a representative view of the challenges
encountered when approximating the ACAS-Xu decision logic with NNs.

\begin{figure}[h!]
    \centering
    \includegraphics[width=11cm]{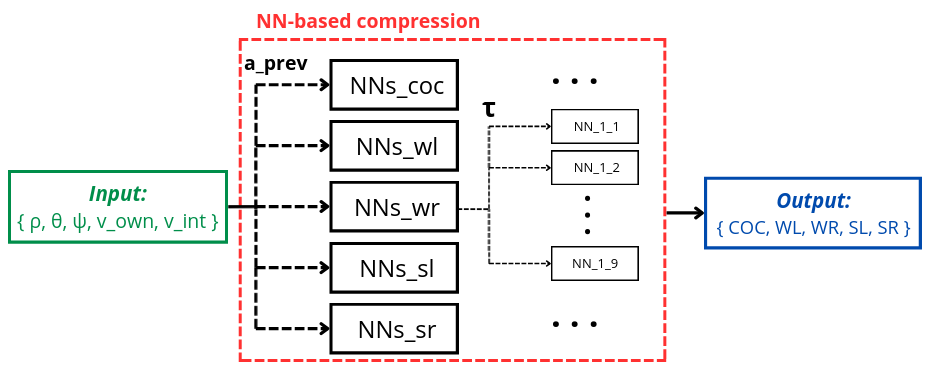}
    \caption{ACAS-Xu Architecture with NN}
    \label{fig:NN}
\end{figure}

The ACAS-Xu advisory decision process is implemented through an array of 45 distinct Deep Neural Networks, each trained to operate within a specific region of the state space.
These networks are obtained by discretizing two ACAS-Xu input dimensions: $\tau$ (time to loss of vertical separation) and $a_{prev}$ (previous advisory).
The architecture overview is shown in Figure \ref{fig:NN}.

Each network shares the same architecture : fully connected topologies with 6 hidden layers of 50 neurons each and ReLU activation functions,
5 inputs corresponding to the remaining ACAS-Xu state variables and 5 outputs representing the advisory scores~\cite{DnnCompression}.

\subsection{NNs analysis}

When trained on LUT data, these models can often reproduce the advisory logic with an average accuracy~\cite{Julian2016ScalableVerification}.
Several structural limitations affect their suitability for safety-critical decision systems.
NNs do not provide deterministic guarantees on their outputs, and small variations in the input space may lead to inconsistent or incorrect advisories \cite{Szegedy2014Adversarial}.
These failures typically occur in localized regions of the state space, where decision boundaries are sensitive, sparse, or difficult to approximate accurately~\cite{Julian2016ScalableVerification,Julian2019PolicyCompression}.

To assess the reliability of these networks, we evaluate a subset of five NNs, arbitrarily selected as the first network associated with each LUT,
The evaluation is performed on two targeted test sets extracted from the original LUTs: 
the COC regions (states for which the LUT returns the advisory Clear of Conflict)
and the Non-COC regions (states requiring an avoidance maneuver like WL, WR, SL or SR). 
For each test set, the evaluation consists in a pointwise comparison between the outputs produced by the neural networks and 
the corresponding LUT advisories over all states contained in the set.

\setlength{\tabcolsep}{4pt}

\begin{center}
    \begin{tabular}{|l|c|c|}
        \hline
        \textbf{NN} & \textbf{LUT Points (COC)} & \textbf{LUT Points (Non-COC)} \\
        \hline
        $NN_{1\_1}$ & 97.76\% & 57.44\% \\
        $NN_{2\_1}$ & 96.58\% & 69.95\% \\
        $NN_{3\_1}$ & 92.30\% & 68.53\% \\
        $NN_{4\_1}$ & 98.54\% & 59.58\% \\
        $NN_{5\_1}$ & 98.22\% & 59.87\% \\
        \hline
    \end{tabular}
\end{center}

Across all tested networks, the results reveal a consistent and marked discrepancy between performance on COC and Non-COC regions.
On COC states, the networks achieve very high conformity rates, typically above 96\%,
indicating that the models are able to accurately learn and reproduce the dominant “no-action” regions of the LUTs.
This behavior is expected, as COC regions occupy a large and relatively homogeneous portion of the state space,
making them easier to approximate using standard supervised learning techniques.
In contrast, performance on Non-COC regions is significantly lower.
Conformity rates for states requiring an avoidance maneuver range from approximately 57\% to 70\%, depending on the network.
These results highlight a systematic weakness of the neural models precisely in the regions where correct behavior is most critical for safety.
Non-COC regions correspond to narrower, more complex, and highly structured decision boundaries, where small changes in the state can trigger different advisories.

The observed performance gap between COC and Non-COC regions highlights a central challenge of neural-network-based LUT compression for ACAS-Xu.
These results indicate that neural approximations alone struggle to provide uniform correctness guarantees across the entire operational domain,
and that average accuracy metrics are insufficient to characterize their reliability in safety-critical settings.
This limitation has motivated the exploration of hybrid solutions that aim to combine the efficiency of neural networks with the exactness of the original LUTs.
In particular, several approaches propose complementing neural decision logic with selected portions of the LUT in regions where the network is known to be unreliable~\cite{SafetyNet,ArthurPhd}.
By reverting to the exact LUT advisory in these cases, such mechanisms, often referred to as Safety Nets,
seek to retain the compactness of neural representations while mitigating their most critical failure modes.
Nevertheless, maintaining such LUT fragments partially offsets the benefits of compression and adds complexity to the overall architecture.


\section{BDD for representing ACAS-Xu logic}
\label{sec:BDD}

The limitations of NNs highlight the need for alternative representations that remain compact
while preserving the exact behavior of the original LUTs. In this context, symbolic data structures offer a promising direction.
Among them, BDDs provide a canonical and deterministic encoding of discrete functions,
making them well suited for representing the ACAS-Xu decision logic without approximation. Building on these properties,
this section introduces our BDD-based compression approach and discusses how it addresses the constraints associated with LUT storage on UAV platforms.

\subsection{Presentation and illustrative example of a BDD}

BDDs are symbolic data structures used to represent Boolean or discrete functions in a compact and canonical form~\cite{Bryant1986}.
A BDD encodes a function as a directed acyclic graph in which each internal node corresponds to the evaluation of a variable,
and each terminal node corresponds to an output value.
Under a fixed variable ordering, the BDD representation is unique,
which provides strong guarantees on determinism and simplifies equivalence checking~\cite{Bryant1986}.
BDDs are widely used in formal verification, model checking, and hardware synthesis due to their ability to represent large state spaces efficiently~\cite{Clarke1999,McMillan1993Symbolic}.

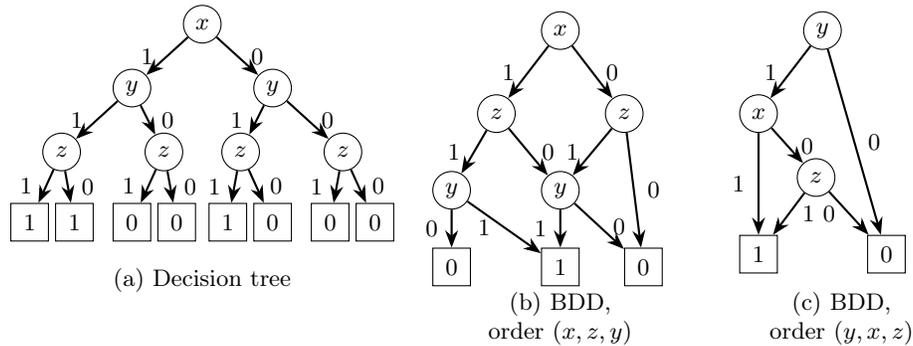
\begin{figure}[!ht]
\centering
\begin{tikzpicture}[
    every node/.style={draw, circle, minimum size=5mm, inner sep=1pt},
    term/.style={draw, rectangle, minimum size=5mm, inner sep=1pt},
    edge/.style={-Stealth, thick},
    scale=0.85
]

\node (x1) at (-8,0) {$x$};
\node (y1) at (-9.1,-1.0) {$y$};
\node (y2) at (-6.9,-1.0) {$y$};
\node (z1) at (-10.2,-2.0) {$z$};
\node (z2) at (-8.6,-2.0) {$z$};
\node (z3) at (-7.4,-2.0) {$z$};
\node (z4) at (-5.8,-2.0) {$z$};

\node[term] (t1) at (-10.7,-3.1) {$1$};
\node[term] (t2) at (-10,-3.1) {$1$};
\node[term] (t3) at (-9.1,-3.1) {$0$};
\node[term] (t4) at (-8.4,-3.1) {$0$};
\node[term] (t5) at (-7.6,-3.1) {$1$};
\node[term] (t6) at (-6.9,-3.1) {$0$};
\node[term] (t7) at (-6,-3.1) {$0$};
\node[term] (t8) at (-5.3,-3.1) {$0$};

\draw[edge] (x1) -- node[left, draw=none] {$1$} (y1);
\draw[edge] (x1) -- node[right, draw=none] {$0$} (y2);

\draw[edge] (y1) -- node[left, draw=none] {$1$} (z1);
\draw[edge] (y1) -- node[right, draw=none] {$0$} (z2);

\draw[edge] (y2) -- node[left, draw=none] {$1$} (z3);
\draw[edge] (y2) -- node[right, draw=none] {$0$} (z4);

\draw[edge] (z1) -- node[left, draw=none] {$1$} (t1);
\draw[edge] (z1) -- node[right, draw=none] {$0$} (t2);

\draw[edge] (z2) -- node[left, draw=none] {$1$} (t3);
\draw[edge] (z2) -- node[right, draw=none] {$0$} (t4);

\draw[edge] (z3) -- node[left, draw=none] {$1$} (t5);
\draw[edge] (z3) -- node[right, draw=none] {$0$} (t6);

\draw[edge] (z4) -- node[left, draw=none] {$1$} (t7);
\draw[edge] (z4) -- node[right, draw=none] {$0$} (t8);

\node[draw=none] at (-8,-4.0) {\small (a) Decision tree};

\node (x2) at (-2.4,-0.1) {$x$};
\node (z5) at (-3.4,-1.4) {$z$};
\node (z6) at (-1.4,-1.4) {$z$};
\node (y3) at (-4.1,-2.6) {$y$};
\node (y4) at (-2.4,-2.6) {$y$};

\node[term] (tb1) at (-4.1,-3.8) {$0$};
\node[term] (tb2) at (-2.4,-3.8) {$1$};
\node[term] (tb3) at (-1.1,-3.8) {$0$};

\draw[edge] (x2) -- node[left, draw=none] {$1$} (z5);
\draw[edge] (x2) -- node[right, draw=none] {$0$} (z6);

\draw[edge] (z5) -- node[left, draw=none] {$1$} (y3);
\draw[edge] (z5) -- node[right, draw=none] {$0$} (y4);

\draw[edge] (z6) -- node[right, draw=none] {$0$} (tb3);
\draw[edge] (z6) -- node[left, draw=none] {$1$} (y4);

\draw[edge] (y3) -- node[left, draw=none] {$1$} (tb2);
\draw[edge] (y3) -- node[left, draw=none] {$0$} (tb1);

\draw[edge] (y4) -- node[left, draw=none] {$1$} (tb2);
\draw[edge] (y4) -- node[right, draw=none] {$0$} (tb3);

\node[draw=none, align=center] at (-2.4,-4.6) {\small (b) BDD, \\ order $(x,z,y)$};

\node (y6) at (1.7,-0.1) {$y$};
\node (x3) at (0.7,-1.4) {$x$};
\node (z7) at (1.6,-2.4) {$z$};

\node[term] (tc4) at (2.7,-3.6) {$0$};
\node[term] (tc1) at (0.7,-3.6) {$1$};

\draw[edge] (y6) -- node[left, draw=none] {$1$} (x3);
\draw[edge] (y6) -- node[right, draw=none] {$0$} (tc4);

\draw[edge] (x3) -- node[right, draw=none] {$0$} (z7);
\draw[edge] (x3) -- node[left, draw=none] {$1$} (tc1);

\draw[edge] (z7) -- node[right, draw=none] {$1$} (tc1);
\draw[edge] (z7) -- node[left, draw=none] {$0$} (tc4);

\node[draw=none, align=center] at (2.0,-4.6) {\small (c) BDD,\\ order $(y,x,z)$};

\end{tikzpicture}
\caption{
Decision tree and BDD representations of $f(x,y,z)$.
}
\label{fig:tree_vs_bdd}
\end{figure}

To illustrate the principle of BDDs, we consider the Boolean function $f(x,y,z) = (x \land y) \lor (y \land z)$
and compare its representation as a decision tree and as BDDs in Figure~\ref{fig:tree_vs_bdd}.

The decision tree in Figure~\ref{fig:tree_vs_bdd}(a) evaluates
function by testing variables sequentially along a root-to-leaf path.
For the function $f(x,y,z)$ this results in multiple distinct subtrees rooted at the variable $y$,
which are reached under different assignments of the preceding
variables $x$ and $z$. A BDD eliminates this redundancy by explicitly sharing equivalent substructures.
Under a fixed variable ordering, identical subfunctions are represented by a single node and reused across multiple evaluation paths.
This effect is illustrated in Figure~\ref{fig:tree_vs_bdd}(b), where the BDD constructed with a suboptimal variable ordering exhibits limited sharing and
therefore remains relatively large.
The importance of the variable ordering becomes apparent when comparing this representation with the BDD obtained under a more suitable ordering.
In Figure~\ref{fig:tree_vs_bdd}(c), placing $y$ early in the ordering exposes common decision patterns, allowing the corresponding subgraphs to be fully shared.
This ability to factorize common decision patterns is essential when representing large discrete decision functions, such as the ACAS-Xu advisory logic,
provided that an appropriate variable ordering is chosen.

\subsection{BDD construction procedure}

In the context of ACAS-Xu, several characteristics of the decision logic make BDDs particularly well suited.
The LUTs map discretized encounter states to a finite set of advisories, forming a deterministic function over a structured and fully discrete domain.
This setting aligns naturally with BDD representations, which can exploit redundancies and common patterns across the input space to
drastically reduce the size of the encoded function.
In our construction, the BDDs capture only the final advisory decision associated with each state,
whereas the original LUTs additionally store the underlying cost or score values used during offline optimization.

\begin{figure}[h!]
    \centering
    \includegraphics[width=9cm]{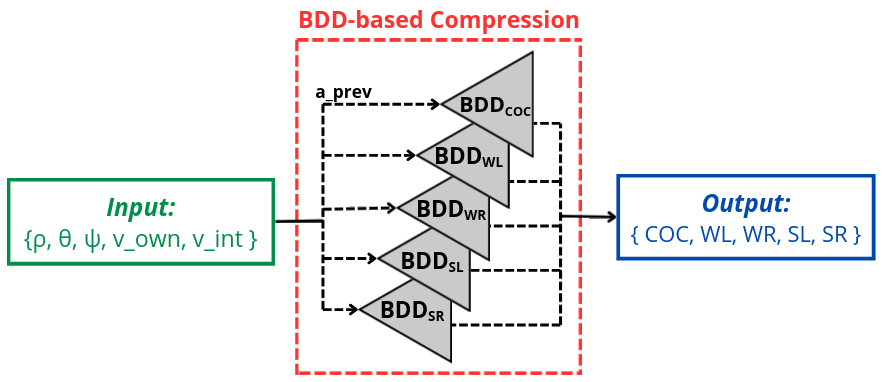}
    \caption{ACAS-Xu Architecture with BDDs}
    \label{fig:Sys_BDD}
\end{figure}

Fundamentally, the decision-making process of the ACAS-Xu is a geometric partitioning problem: large regions of the state space share the same optimal advisory,
leading to significant data redundancy. By representing these LUTs as BDDs, we can exploit this underlying structure.
Moreover, BDDs preserve the exact behavior of the original LUTs: no approximation is introduced,
and every state maps deterministically to the same advisory as in the reference tables.
At the end, our idea is to have 5 BDDs, corresponding to the 5 LUTs of
the systems, depending on the previous action, cf.~Figure~\ref{fig:Sys_BDD}.

In our approach, BDDs are used not merely as a symbolic encoding of the ACAS-Xu decision logic, but explicitly as a classifier over the discretized state space.
Instead of representing a single Boolean function, the diagram encodes the advisory decision by associating a Boolean predicate to each of the five possible actions.
Each predicate characterizes precisely the set of states for which the LUT prescribes that advisory.
This structure allows the system to determine the prescribed action by evaluating which advisory predicate evaluates to \texttt{true} for a given input state.
Because the advisory sets are provably disjoint, exactly one predicate will hold for any valid encounter state, providing a deterministic and unambiguous classification mechanism.
Viewed this way, the BDD functions are an exact, symbolic multi-class classifier whose decision boundaries coincide perfectly with those of the original ACAS-Xu LUTs.

To encode the ACAS-Xu LUTs as BDDs, each LUT entry is first identified by a discrete state tuple $(\tau, \rho, \theta, \psi, v_{own}, v_{int})$
where each component corresponds to a predefined quantization level of the associated state variable.
Each state is therefore mapped to a tuple of integer indices $(\tau^{i}, \rho^{i}, \theta^{i}, \psi^{i}, v_{own}^{i}, v_{int}^{i})$
where each index uniquely identifies the corresponding table variable along that dimension.
These integer indices are then encoded in binary to form the Boolean input variables of the BDD.
For each dimension, we select the minimum number of bits required to represent all possible index values.
Specifically, the three parameters $\tau$, $v_{own}$ and $v_{int}$ are encoded each on four bits,
in order to cover the 10 possible values of $\tau$, the 12 values of $v_{own}$, and the 12 values of $v_{int}$.
Similarly, six bits are used to encode each of the three last parameters $\rho$, $\theta$ and $\psi$, in order to cover the 41 values of $\rho$,
the 39 values of $\theta$, and the 39 values of $\psi$.
This encoding yields a compact Boolean representation of the discrete ACAS-Xu state space, suitable for symbolic manipulation using BDDs.

Because the ACAS-Xu decision problem is fundamentally geometric, the structure of the advisory logic exhibits strong spatial regularities.
Advisories tend to form contiguous regions in the discretized state space, corresponding to avoidance maneuvers that vary smoothly with respect to the state parameters.
As a result, it is unlikely for the recommended action to change abruptly between two successive index values along any single dimension.
We have already emphasized the critical role of variable ordering in achieving compact BDD representations.
In the same spirit, the choice of how discrete indices are encoded at the bit level has a direct impact on the amount of structural sharing that can be exploited during BDD construction.
To promote compactness, we therefore encode all state indices using fixed-size Gray code representations.
This encoding ensures that consecutive index values differ by only a single bit, thereby preserving the geometric proximity of neighboring states in the Boolean domain~\cite{Doran2007GrayCode}.
By limiting the number of bit flips between adjacent discrete states, Gray encoding increases the likelihood that nearby regions of the state space share common subgraphs in the BDD.
This, in turn, facilitates the merging of equivalent decision structures and leads to more compact diagrams.

Each discretized state is then mapped to a single Boolean cube, defined as the conjunction of all bit literals corresponding to that state.
This cube evaluates to \texttt{true} for that exact state and for no other. For each advisory, we collect all corresponding cubes and merge them through disjunction to
form one BDD root per advisory. This produces five separate BDDs, one for each LUT action, that exactly encode the content of the table,
since each LUT entry corresponds to a unique cube in its associated root.

\subsection{BDD implementation}

Our construction relies on the CUDD library~\cite{Somenzi1998CUDD}, accessed through the \texttt{dd.cudd} Python wrapper~\cite{ddPython}.
CUDD provides an optimized BDD manager that supports fast Boolean operations, unique-table-based node canonicalization, garbage collection,
and dynamic variable reordering. These capabilities are essential as the ACAS-Xu LUTs contain millions of entries.

The construction of the BDDs is performed incrementally.
All Boolean variables are declared at initialization and assigned an initial fixed ordering that directly reflects the structure of the ACAS-Xu LUTs.
This ordering, while convenient and intuitive, is an arbitrary design choice.
Specifically, variables are ordered dimension by dimension and bit by bit, following the natural LUT layout, where index $0$ denotes the most significant bit of each encoded parameter:
\[
\left[
\begin{array}{l}
  \tau^0, \tau^1, \tau^2, \tau^3, \rho^0, \rho^1, \rho^2, \rho^3, \rho^4, \rho^5, \theta^0, \theta^1, \theta^2, \theta^3, \theta^4, \theta^5, \psi^0, \psi^1, \psi^2, \psi^3, \psi^4, \psi^5, \\
  v_{own}^{0}, v_{own}^{1}, v_{own}^{2}, v_{own}^{3}, v_{intr}^{0}, v_{intr}^{1}, v_{intr}^{2}, v_{intr}^{3}
\end{array}
\right]
\]

The LUT files are processed in chunks to limit peak memory usage.
Each chunk is translated into a set of Boolean cubes, merged into a temporary BDD, and then incorporated into the corresponding advisory root.
During construction, we periodically invoke CUDD’s dynamic variable reordering mechanisms~\cite{Rudell1993Sifting},
which aim to identify a variable order that minimizes the overall diagram size.
The impact of variable reordering is known to be substantial.
For example, when constructing the BDD corresponding to the LUT with previous advisory SR,
the initial ordering yields a diagram containing $902,180$ nodes.
After applying CUDD’s reordering, the diagram size is reduced to $483,638$ nodes, representing a reduction of nearly $46\%$.
The resulting variable order is:
\[
\left[
\begin{array}{l}
\rho^0,\rho^1,\psi^1,\psi^0,\theta^0,\theta^1, v_{int}^{1},v_{int}^{0},v_{own}^{0},v_{own}^{1}, \theta^2,\psi^2,\psi^3,\theta^3,\theta^4,\psi^4,\theta^5,\psi^5, \\
v_{int}^{2},v_{own}^{2},v_{int}^{3},v_{own}^{3}, \rho^4,\rho^5,\rho^3, \tau^3,\rho^2,\tau^2,\tau^0,\tau^1
\end{array}
\right]
\]
An analysis of the variable ordering given by CUDD provides additional insight into the structure of the ACAS-Xu decision logic.
In particular, the bits associated with the variable $\tau$ are consistently pushed to the end of the ordering,
indicating that $\tau$ has a limited influence on the advisory outcome compared to the other state variables.
By contrast, the remaining dimensions are ordered such that their most significant bits appear early in the diagram.
This observation suggests that the partitioning of neural networks along the $\tau$ dimension, as performed in~\cite{DnnCompression},
may not be the most relevant decomposition strategy.
Finally, when initializing CUDD’s dynamic reordering with the ordering obtained here, the algorithm converges to an almost identical configuration,
which indicates that this ordering seems closed to an optimal solution.

Each advisory root is dumped independently during construction, ensuring debuggability and reproducibility.
And then final diagrams are stored and loaded using the DDDMP format provided by CUDD.

\subsection{Partitioning check and root reduction}

Once the five advisory BDDs are constructed, we verify that they form a partition of the discretized state space.
First, we check that they are pairwise disjoint by computing the conjunction of each pair of BDDs. In all our experiments,
every such conjunction evaluates to \texttt{False}, confirming that the advisory sets are mutually exclusive, as expected from the LUT semantics.
Second, we verify that their union covers the entire state space:
$root_{1} \lor root_{2} \lor root_{3} \lor root_{4} \lor root_{5} = \top$

Together, these two properties ensure a porper partitioning: every discretized state belongs to exactly one advisory set.
This partitioning property has an important practical consequence. Since the advisory sets are mutually exclusive and collectively exhaustive,
one of the BDDs can be represented implicitly as the complement of the union of the others. In particular, we remove the largest BDD and define it as
$root_{5} = \lnot(root_{1} \lor root_{2} \lor root_{3} \lor root_{4})$

When the eliminated BDD is substantially larger than the others, this substitution leads to a noticeably more compact overall representation.
In our construction, we therefore retain only four explicit BDDs and rely on this complement-based representation for the fifth advisory.

\subsection{Final single-root BDD}

The advisory-specific BDDs are assembled into a single, deployable diagram by introducing two selector variables, $bdd0$ and $bdd1$.
Each advisory is associated with a fixed pattern over these two variables. For example, in the context of the LUT corresponding to the previous action SR,
we have in our BDD encoding:

\begin{figure}[h!]
    \centering
    \begin{minipage}{0.45\textwidth}
        \[
        \begin{array}{lcl}
        sel_{COC} &=& bdd1 \land bdd0 \\
        sel_{SL}  &=& \lnot bdd1 \land bdd0 \\
        sel_{WR}  &=& bdd1 \land \lnot bdd0 \\
        sel_{SR}  &=& \lnot bdd1 \land \lnot bdd0
        \end{array}
        \]
    \end{minipage}
    \hfill
    \begin{minipage}{0.54\textwidth}
        \centering
        \includegraphics[width=6.5cm]{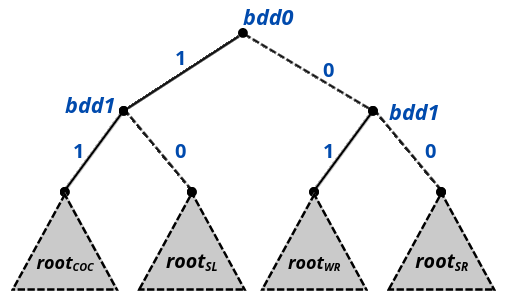}
        \caption{Final Single-Root $BDD_{SR}$ illustration ($a_{prev}$ = SR)}
        \label{fig:BDD}
    \end{minipage}
\end{figure}

For each retained advisory, we construct a guarded subgraph of the form $sel_{a} \land root_{a}$ and combine all guarded subgraphs through disjunction:
$$
global_{root} = \bigvee_{a \in Advisories} \left( sel_{a} \land root_{a} \right)
$$

The two selector variables $bdd0$ and $bdd1$ are introduced as additional Boolean inputs and are placed at the very top of the global BDD.
Their position in the variable ordering is fixed and they are explicitly excluded from the dynamic reordering process.
As a result, every evaluation of the global diagram begins by assigning values to these two selector variables,
thereby deterministically selecting the advisory of interest before any state-dependent decision logic is traversed.
By fixing the selector variables at the root of the diagram, the structure of each advisory-specific subgraph remains unchanged and fully isolated from the others.
Moreover, it simplifies the use of the BDD in deployment: selecting and testing a particular advisory reduces to setting $(bdd1,bdd0)$
to the corresponding binary pattern and evaluating the resulting Boolean function.

Since the advisory roots are disjoint, each input state satisfies exactly one predicate.
The fifth advisory is implicitly represented by the negation of the union of the four explicit ones.
Let us notice that the four retained advisories may differ from one LUT to another. In the case of the LUT corresponding to the previous action SR,
the four retained advisories are COC, SL, WR and SR.
We show in Figure \ref{fig:BDD} the final single-root BDD in the case of $a_{prev} = SR$.


\section{Properties verification using BDD}
\label{sec:verif}


\subsection{Verification through domain-specific proporties}
\label{sec:reluplex_verif}

The BDDs obtained were first validated against the original ACAS-Xu LUTs.
As expected, the symbolic representation ensures strict equivalence:
\emph{for every discretized state contained in the LUT domain, the advisory returned by the BDD-based decision logic matches the reference LUT output.}

Beyond functional validation, a key advantage of the BDD-based representation is that it naturally enables formal verification of decision logic properties.
Because both the system behavior and the verification conditions can be expressed symbolically,
property checking reduces to standard boolean reasoning over decision diagrams.

Following the methodology introduced in prior work such as Reluplex-based analyses \cite{Reluplex}, domain-specific
properties can be formalized as constraints over the encounter state and the issued advisory. In our setting, the properties are encoded
by two separate BDDs: $BDD_{prop}$, which characterizes the set of
states for which the property’s preconditions hold and $R$ which
encodes the expected outcome of the property.
The compressed ACAS-Xu logic itself is represented by $BDD_{acas}$.
Property satisfaction can then be checked symbolically by verifying the logical implication: $BDD_{acas} \land BDD_{prop} \Rightarrow R$.
This is equivalently tested by checking the emptiness of the following intersection: $BDD_{acas} \land BDD_{prop} \land \lnot R$.
If this intersection evaluates to the empty BDD, the property is formally proven to hold for all states satisfying the constraints.
Conversely, if the intersection is non-empty, the satisfying assignments correspond to concrete counterexamples that witness a violation of the property.
These counterexamples can be extracted directly from the BDD, providing precise and interpretable diagnosis.

The results of our verification experiments are reported on Table~\ref{tab:BDD_properties_tests}.
These properties were originally proposed in \cite{Reluplex} and verified on a small subset of the NNs,
but not on all 45 networks representing the complete system without a clear explanation.
Extending the evaluation to the complete ACAS-Xu domain may reveal unforeseen behaviors in certain regions of the state space.

\setlength{\tabcolsep}{6pt}

\begin{table}[h!]
    \centering
    \caption{Verification of domain-specific properties from \cite{Reluplex} on BDDs.
            Property $P_8$ applies only to scenarios with $a_{prev} = WL$ and can therefore be evaluated only on $BDD_{WL}$.}
    \label{tab:BDD_properties_tests}
    \begin{tabular}{|l|l|l|l|l|l|}
        \hline
        \textbf{P} & \textbf{BDD}\textsubscript{\textbf{COC}} & \textbf{BDD}\textsubscript{\textbf{WL}} & \textbf{BDD}\textsubscript{\textbf{WR}} & \textbf{BDD}\textsubscript{\textbf{SL}} & \textbf{BDD}\textsubscript{\textbf{SR}} \\
        \hline
        $\mathbf{P}_{\mathbf{1}}$ & \texttt{\textcolor{red}{Invalid}} & \texttt{\textcolor{green}{Valid}} & \texttt{\textcolor{green}{Valid}} & \texttt{\textcolor{red}{Invalid}} & \texttt{\textcolor{red}{Invalid}} \\
        $\mathbf{P}_{\mathbf{2}}$ & \texttt{\textcolor{red}{Invalid}} & \texttt{\textcolor{green}{Valid}} & \texttt{\textcolor{green}{Valid}} & \texttt{\textcolor{red}{Invalid}} & \texttt{\textcolor{red}{Invalid}} \\
        $\mathbf{P}_{\mathbf{3}}$ & \texttt{\textcolor{green}{Valid}} & \texttt{\textcolor{green}{Valid}} & \texttt{\textcolor{green}{Valid}} & \texttt{\textcolor{green}{Valid}} & \texttt{\textcolor{green}{Valid}} \\
        $\mathbf{P}_{\mathbf{4}}$ & \texttt{\textcolor{green}{Valid}} & \texttt{\textcolor{green}{Valid}} & \texttt{\textcolor{green}{Valid}} & \texttt{\textcolor{green}{Valid}} & \texttt{\textcolor{green}{Valid}} \\
        $\mathbf{P}_{\mathbf{5}}$ & \texttt{\textcolor{green}{Valid}} & \texttt{\textcolor{green}{Valid}} & \texttt{\textcolor{red}{Invalid}} & \texttt{\textcolor{green}{Valid}} & \texttt{\textcolor{red}{Invalid}} \\
        $\mathbf{P}_{\mathbf{6}}$ & \texttt{\textcolor{green}{Valid}} & \texttt{\textcolor{green}{Valid}} & \texttt{\textcolor{green}{Valid}} & \texttt{\textcolor{green}{Valid}} & \texttt{\textcolor{green}{Valid}} \\
        $\mathbf{P}_{\mathbf{7}}$ & \texttt{\textcolor{green}{Valid}} & \texttt{\textcolor{red}{Invalid}} & \texttt{\textcolor{red}{Invalid}} & \texttt{\textcolor{green}{Valid}} & \texttt{\textcolor{green}{Valid}} \\
        $\mathbf{P}_{\mathbf{8}}$ & N/A & \texttt{\textcolor{green}{Valid}} & N/A & N/A & N/A \\
        $\mathbf{P}_{\mathbf{9}}$ & \texttt{\textcolor{green}{Valid}} & \texttt{\textcolor{green}{Valid}} & \texttt{\textcolor{green}{Valid}} & \texttt{\textcolor{green}{Valid}} & \texttt{\textcolor{green}{Valid}} \\
        $\mathbf{P}_{\mathbf{10}}$ & \texttt{\textcolor{green}{Valid}} & \texttt{\textcolor{green}{Valid}} & \texttt{\textcolor{green}{Valid}} & \texttt{\textcolor{green}{Valid}} & \texttt{\textcolor{green}{Valid}} \\
        \hline
    \end{tabular}
\end{table}

The results reported in Table~\ref{tab:BDD_properties_tests} admit a twofold interpretation.
First, for each \texttt{Invalid} property, the BDD-based verification procedure produces an explicit counterexample in the form of a concrete discretized encounter state.
Because the constructed BDDs are exact symbolic representations of the original ACAS-Xu LUTs,
these counterexamples can be directly evaluated against the LUTs themselves.
In all cases, the counterexamples identified by the BDD verification process correspond to genuine violations in the reference LUTs.
These results confirm the correctness of the BDD encoding and reveal a discrepancy between properties verified on NN approximations and 
the behavior encoded in the original ACAS-Xu LUTs.
Specifically, several domain-specific properties that were verified on a limited set of neural networks using Reluplex~\cite{Reluplex}
do not hold when evaluated against the LUTs.
This observation raises a broader question about how such properties should be interpreted and validated.
The second concerns the effectiveness of BDDs as a verification substrate and is detailed in~\ref{sec:BDD_for_tool}.

\subsection{BDDs as an effective verification tool for ACAS-Xu}
\label{sec:BDD_for_tool}



In contrast to LUT-based verification, which is computationally prohibitive, and NN-based approaches,
which require specialized solvers and introduce approximation artifacts, BDDs offer a unique combination of exactness and tractability.
Properties can be encoded directly as Boolean constraints and composed with the system BDD using standard logical operations,
enabling efficient validation and precise counterexample extraction at the scale of the full decision logic.

Moreover, the symbolic nature of BDDs significantly broadens the class of properties that can be verified.
While verification of neural-network approximations typically relies on interval-based abstractions over input parameters,
BDDs naturally support relational properties involving multiple state variables.
This makes possible to reason about more expressive operational and safety constraints without modifying the underlying system representation,
and supports incremental verification workflows well suited for certification and safety analysis.
We provide below an example of such a relational operational property that can be easily verified using the BDD-based framework.

\setcounter{property}{10}
\begin{property}
    If the intruder is approaching the ownship from behind at a lower speed, the network will advise COC. \\
    Input constraints : \textbf{$v_{int} \leq v_{own}$}, $8500.0 \leq \rho \leq 62000.0$, -$3.1416 \leq \theta \leq $-$3.1416 + 0.01$, -$0.06 \leq \psi \leq 0.06$ \\
    Desired Output : COC
\end{property}

The verification of such a property using BDDs reduces also to a simple symbolic intersection test.
First, the input constraints are encoded as a BDD, denoted $BDD_{C}$.
This BDD is constructed as the conjunction of individual BDDs encoding each interval constraint on the state variables,
together with an additional BDD capturing the relational constraint between the two speed variables.
The property holds if and only if there exists no state satisfying the constraints for which the system issues a non-COC advisory.
Formally, this amounts to checking the emptiness of the following intersection: $ BDD_{C} \wedge BDD_{acas} \wedge \neg COC$.
If this intersection is empty, the property is validated; otherwise, any satisfying assignment provides a concrete counterexample.
The results of this verification are reported below.
It illustrates how complex domain-specific and relational operational properties can be expressed and verified efficiently within the BDD framework.

\setlength{\tabcolsep}{5pt}

\begin{center}
    \begin{tabular}{|l|l|l|l|l|l|}
        \hline
        \textbf{P} & \textbf{BDD}\textsubscript{\textbf{COC}} & \textbf{BDD}\textsubscript{\textbf{WL}} & \textbf{BDD}\textsubscript{\textbf{WR}} & \textbf{BDD}\textsubscript{\textbf{SL}} & \textbf{BDD}\textsubscript{\textbf{SR}} \\
        \hline
        $\mathbf{P}_{\mathbf{11}}$ & \textcolor{green}{Valid} & \textcolor{green}{Valid} & \textcolor{green}{Valid} & \textcolor{green}{Valid} & \textcolor{green}{Valid} \\
        \hline
    \end{tabular}
\end{center}

\setlength{\textfloatsep}{8pt}
\setlength{\floatsep}{6pt}
\setlength{\intextsep}{8pt}

\section{Onboard deployment}
\label{sec:expe}

\subsection{\texttt{C} code generation}
\label{sec:C_generation}

Using the proposed construction strategy, we obtain five BDDs corresponding to the five ACAS-Xu LUTs used for horizontal collision avoidance, 
stored in the standard DDDMP format.
These DDDMP representations are then parsed to generate an equivalent standalone \texttt{C} implementation of the BDD evaluation logic,
in which the diagram structure and variable ordering are explicitly encoded without relying on any external libraries.
The resulting implementation consists of approximately $1.5$ million lines of \texttt{C} code encoding the BDD structure, 
together with a lightweight runtime of a few hundred lines implementing the evaluation procedure.
This \texttt{C}-based implementation significantly improves embeddability, enabling deployment on a wide range of embedded platforms.
Finally, correctness is ensured by exhaustively validating the generated code against all entries of the corresponding LUT, 
confirming a faithful transfer from the symbolic representation to executable logic.

\subsection{Memory footprint}

Beyond deployment, our objective is to systematically compare the proposed BDD based compression with NN based approaches.
BDDs provide substantial reductions in storage requirements by exploiting structural regularities in the decision logic.
In practice, the storage cost of a BDD-based implementation is determined by the size of the generated \texttt{C} code required to evaluate the diagram.
We measured then the memory footprint of the \texttt{C} code compiled with \texttt{gcc} and the option \texttt{-O3}.
This code fully captures the decision logic and can be directly deployed on embedded targets without requiring additional data structures or lookup tables.
In comparaison, the memory footprint of the neural-network-based implementation can be directly estimated from the network architecture
\cite{DnnCompression}.
The raw size of a single network is approximately 53~KB, leading to a total size of about
$9 * 53$ KB, i.e., roughly 0.5~MB, to represent one LUT. However, the additional memory required by the Safety Net mechanism,
which is necessary to guarantee correctness when neural networks alone are insufficient \cite{SafetyNet,Unsafe}, is missing.
According to \cite{ArthurPhd}, once the Safety Net is included,
the overall storage requirement grows to approximately 65 MB in a setting where one LUT parameter is fixed.

\setlength{\tabcolsep}{8pt}

\begin{table}[h!]
    \centering
    \caption{Memory footprint comparison of ACAS-Xu representations (in MB)}
    \label{tab:memory_footprint_comparison}
    \begin{tabular}{|l|c|c|c|c|c|}
        \hline
        \textbf{} & \textbf{COC} & \textbf{WL} & \textbf{WR} & \textbf{SL} & \textbf{SR} \\
        \hline
        \textbf{LUT} & 944 & 944 & 944 & 944 & 944 \\
        \hline
        \textbf{NN} & 0.5 & 0.5 & 0.5 & 0.5 & 0.5 \\
        \hline
        \textbf{NN + Safety Net} & $\ge$ 65 & $\ge$ 65 & $\ge$ 65 & $\ge$ 65 & $\ge$ 65  \\
        \hline
        \textbf{BDD} & 2.7 & 1.8 & 1.8 & 2.8 & 2.8 \\
        \hline
    \end{tabular}
\end{table}

Table~\ref{tab:memory_footprint_comparison} highlights the drastic reduction in storage requirements achieved by BDDs compared to the original LUTs, while preserving exact behavior.
It is important to note that LUTs store the scores for all four possible actions for each state, whereas BDDs encode the decision logic without explicitly storing all these scores.
Thus, to make a fair comparison of memory footprint per action, the effective memory requirement of the LUTs would need to be divided by 10: roughly 94~MB per action,
it remains substantially larger than the memory required by BDDs.
Although neural networks achieve even smaller raw sizes, the additional Safety Net required to ensure correctness largely offsets this benefit.

\subsection{Execution time}

Finally, BDD evaluation incurs significantly lower and more predictable execution times than alternative representations~\cite{Drechsler1998}.
We conducted a comprehensive comparative study of three alternative implementations of the ACAS-Xu horizontal collision avoidance logic:
the original LUTs, NN-based approximations, and the proposed BDD-based representation.
This study was performed across multiple platforms, ranging from a standard desktop workstation to embedded CPU boards and an FPGA-based implementation,
in order to assess execution time and deployability under representative operational constraints.

All reported execution-time measurements follow a common experimental protocol.
Each experiment consists of evaluating the corresponding decision logic on $10{,}000{,}000$ randomly generated encounter states,
representing slightly more than $10\%$ of the states contained in a single LUT.
Moreover, performance evaluation focuses exclusively on the BDD constructed from the SR LUT, which is the largest among the four retained advisory BDDs.
All measurements are therefore conducted in the context of a previous advisory $a_{prev} = SR$,
placing the evaluation in a worst-case configuration for the BDD-based approach and providing a conservative assessment of its execution performance.

\subsubsection{Desktop evaluation:} 
The LUT-, NN-, and BDD-based systems were executed on a Linux workstation equipped with an Intel Core i7 processor.
The BDD-based system corresponds to the standalone \texttt{C} code generated from the DDDMP representation, as described in Section~\ref{sec:C_generation}.
The NN-based implementation relies on a \texttt{C++} program using the NNet inference library~\cite{sisl_nnet_2025},
which loads the appropriate network stored in the \texttt{.nnet} format.
Finally, the LUT-based system consists of a \texttt{C} program that loads the original ACAS-Xu LUT from its binary representation and performs direct table lookups.
This desktop experiment provides a controlled reference point for comparing execution latency and variability across the three representations.
The results, reported in Table~\ref{tab:desktop_results}, show that the BDD-based implementation achieves the lowest average execution time.
More importantly, it also exhibits the smallest worst-case execution time, a critical property for safety-critical real-time systems.

\setlength{\tabcolsep}{8pt}

\begin{table}[h!]
    \centering
    \caption{Time execution on desktop ($\mu s$)}
    \label{tab:desktop_results}
    \begin{tabular}{|l|S|S|S|}
        \hline
         & {\textbf{t}\textsubscript{\textbf{min}}} & {\textbf{t}\textsubscript{\textbf{max}}} & {\textbf{t}\textsubscript{\textbf{moy}}} \\
        \hline
        \textbf{LUT} & 0.970 & 7963.917 & 123.348 \\
        \hline
        \textbf{NN} & 0.009 & 3010.368 & 1.584 \\
        \hline
        \textbf{BDD} & 0.087 & 2640.453 & 0.453 \\
        \hline
    \end{tabular}
\end{table}

\subsubsection{Embedded CPU deployment:}
For both the BDD- and NN-based systems, the same compilation and execution pipelines were reused across a range of embedded CPU platforms
to evaluate the execution performance.
Experiments were conducted on several embedded boards covering diverse architectures and operating environments,
including the BeagleBone Green~\cite{BeagleBoneGreen2026},
the BeagleBone X15~\cite{BeagleBoneX15_2026},
a RISC-V development board based on the StarFive JH7110~\cite{StarFiveJH7110_2026},
and the Kria KR260 platform~\cite{KriaKR260_2026}.
Due to their prohibitive memory footprint, the original ACAS-Xu LUTs could not be loaded on these embedded systems and were therefore excluded from this stage of the evaluation.
Nevertheless, the comparison between BDDs and neural networks across heterogeneous embedded CPUs provides a clear indication of their relative execution performance, predictability, and scalability.
The corresponding results are summarized in Tables~\ref{tab:embedded_cpu_results_1}, Tables~\ref{tab:embedded_cpu_results_2}, 
Tables~\ref{tab:embedded_cpu_results_3} and Tables~\ref{tab:embedded_cpu_results_4}.

\setlength{\tabcolsep}{3pt}

\begin{table}[h!]
\centering
\caption*{Time execution on embedded platforms ($\mu s$)}

\begin{minipage}{0.48\textwidth}
\centering
\caption{BEAGLEBONE Green}
\label{tab:embedded_cpu_results_1}
\begin{tabular}{|l|S|S|S|}
\hline
 & {\textbf{t}\textsubscript{\textbf{min}}} & {\textbf{t}\textsubscript{\textbf{max}}} & {\textbf{t}\textsubscript{\textbf{moy}}} \\
\hline
\textbf{NN} & 1.125 & 16358.000 & 514.137 \\
\textbf{BDD} & 3.125 & 5785.916 & 7.384 \\
\hline
\end{tabular}
\end{minipage}
\hfill
\begin{minipage}{0.48\textwidth}
\centering
\caption{BEAGLEBONE X-15}
\label{tab:embedded_cpu_results_2}
\begin{tabular}{|l|S|S|S|}
\hline
 & {\textbf{t}\textsubscript{\textbf{min}}} & {\textbf{t}\textsubscript{\textbf{max}}} & {\textbf{t}\textsubscript{\textbf{moy}}} \\
\hline
\textbf{NN} & 0.487 & 5267.309 & 71.611 \\
\textbf{BDD} & 1.626 & 1064.002 & 2.860 \\
\hline
\end{tabular}
\end{minipage}

\vspace{0.5cm}

\begin{minipage}{0.48\textwidth}
\centering
\caption{RISC-V}
\label{tab:embedded_cpu_results_3}
\begin{tabular}{|l|S|S|S|}
\hline
& {\textbf{t}\textsubscript{\textbf{min}}} & {\textbf{t}\textsubscript{\textbf{max}}} & {\textbf{t}\textsubscript{\textbf{moy}}} \\
\hline
\textbf{NN} & 0.500 & 4574.000 & 278.890 \\
\textbf{BDD} & 3.000 & 404.250 & 5.396 \\
\hline
\end{tabular}
\end{minipage}
\hfill
\begin{minipage}{0.48\textwidth}
\centering
\caption{KRIA KR260}
\label{tab:embedded_cpu_results_4}
\begin{tabular}{|l|S|S|S|}
\hline
& {\textbf{t}\textsubscript{\textbf{min}}} & {\textbf{t}\textsubscript{\textbf{max}}} & {\textbf{t}\textsubscript{\textbf{moy}}} \\
\hline
\textbf{NN} & 0.090 & 1400.424 & 198.899 \\
\textbf{BDD} & 1.690 & 188.032 & 3.136 \\
\hline
\end{tabular}
\end{minipage}

\end{table}

Across all tested boards, BDD evaluation exhibits consistently low and predictable execution times, on the order of a few microseconds per query.
In contrast, neural-network execution times are one to two orders of magnitude larger on average and exhibit significantly higher variability,
reflecting both computational cost and memory-access overheads.

\subsection{FPGA deployment}

Results obtained from the onboard deployment of BDD on embedded CPU-based boards indicate that the execution times are generally low and compatible with integration into high-rate real-time systems. 
The worst-case execution time can remain significantly higher than the global average. To reduce the CPU load and increase our BDD evaluation logic efficiency, we investigated the acceleration of BDD on a FPGA-based  Xilinx ZCU102 MPSoC architecture~\cite{xilinx_ultrascale_architecture_datasheet_2022}. 
This board contains a large FPGA device with up to 32 Mb of Block Random-Access Memory (BRAM, on-chip memory), separated in 912 blocks of 36 Kb each. The FPGA also includes a lot of Digital Signal Processors (DSPs) and Configurable Logic Blocks (CLBs).

Using Vitis High-Level Synthesis (HLS)~\cite{vitis_hls_2025}, our generated \texttt{C} code 
(detailled in Section \ref{sec:C_generation}) was adapted to meet Xilinx design constraints, 
in particular the AXI (Advanced eXtensible Interface) register-based interface. 
A custom BDD IP core was subsequently generated and integrated into the Board Support Package (BSP) of the Xilinx ZCU102 evaluation board. 
Input and output data were exchanged between a bare-metal \texttt{C} application and the BDD IP core via memory-mapped AXI4-Lite registers. 
The IP core exhibits very low logic utilization, requiring less than 2\% of the available CLBs and no DSP blocks, 
while its memory usage is significantly higher, consuming more than 70\% of the available BRAM resources. The timing measurements, 
including AXI register read and write operations, demonstrate excellent performance, with very low and highly stable processing times 
(see Table~\ref{tab:fpga_time_results}). Overall, these results indicate that, 
despite the relatively high memory usage inherent to BDD structure that must be considered for deployment, 
employing an FPGA as a BDD accelerator can be a highly efficient solution for the ACAS-Xu decision logic.

\setlength{\tabcolsep}{8pt}

\begin{table}[h!]
    \centering
    \caption{Time execution of the BDD IP core on ZCU 102 FPGA ($\mu s$)}
    \label{tab:fpga_time_results}
    \begin{tabular}{|l|l|l|l|}
        \hline
        \textbf{} & \textbf{t}\textsubscript{\textbf{min}} & \textbf{t}\textsubscript{\textbf{max}} & \textbf{t}\textsubscript{\textbf{moy}} \\
        \hline
        \textbf{BDD} & 1.200 & 2.180 & 1.408 \\
        \hline
    \end{tabular}
\end{table}

\section{Conclusion}
\label{sec:conclusion}

This paper introduced a new approach for compressing the operational ACAS-Xu horizontal collision avoidance logic based on BDDs.
Unlike approximation-based techniques, the proposed method provides an exact symbolic encoding of the certified ACAS-Xu LUTs:
every discretized encounter state is mapped to the same advisory as in the reference LUTs, without loss of fidelity.
Beyond compression, we demonstrated through extensive experimentation that this symbolic representation enables efficient and predictable execution on a wide range of platforms,
from desktop CPUs to embedded ARM, RISC-V, and FPGA-based systems.
Across all tested configurations, the BDD-based implementation consistently exhibits lower and more stable execution times than both LUT- and NN-based solutions,
making it particularly attractive for safety-critical and energy-constrained embedded deployments.

A second major contribution of this work is the use of BDDs as a formal verification interface for ACAS-Xu LUTs.
Thanks to the exact correspondence between the LUTs and their BDD encodings, verifying a property on the BDD is equivalent to verifying it directly on the ground-truth decision logic.
This provides a practical symbolic framework for reasoning about the full-scale ACAS-Xu behavior without relying on approximations or surrogate models.
Within this framework, both the system behavior and domain-specific operational properties can be expressed uniformly as Boolean constraints,
reducing verification to conjunction and emptiness checks.
Beyond this unification, the simplicity and expressiveness of the BDD-based formalism make it possible to efficiently verify not only interval-based properties,
but also more expressive relational properties involving interactions between multiple state variables.
This significantly expands the class of properties that can be analyzed compared to NN-based verification approaches,
which typically rely on restrictive abstractions and simple box-shaped input domains.
As a result, the BDD framework enables domain experts, such as flight dynamics or operational safety specialists,
to express more meaningful operational properties that closely reflect real-world aircraft behavior, while remaining amenable to exact and automated verification.
At the same time, this verification capability also raises a fundamental modeling question.
Several operational properties previously shown to hold on NN approximations are violated by the LUTs themselves,
despite the latter being the reference of the ACAS-Xu system.
This discrepancy prompts a deeper discussion on where the ground truth of the system should reside:
in the exact decision logic encoded by the LUTs, or in higher-level operational specifications intended to capture physical or operational expectations.

Despite its advantages, the current approach has a clear limitation.
The constructed BDDs encode only the final advisory decision and do not preserve the action scores stored in the original LUTs.
As a consequence, the proposed representation correctly supports single-intruder scenarios but cannot directly handle multi-intruder encounters,
where ACAS-Xu selects advisories based on a minimum-over-costs aggregation.
Extending the BDD framework to incorporate cost information remains an open challenge.
A straightforward encoding of action scores would however lead to a substantial increase in BDD size, 
as encoding minimum cost values would require approximately 15 additional bits.
This observation points towards future research directions aimed at developing more compact symbolic representations for quantitative decision logic.
A potential direction to address this issue is to further reduce the BDD size by exploiting regions of the LUT where the advisory is constant.
Large portions of the ACAS-Xu state space correspond to trivial or unambiguous situations (e.g., COC regions),
which could be statically identified and handled separately before BDD evaluation.
In BDD terms, this amounts to defining "don’t-care sets" over such regions and simplifying the decision function accordingly~\cite{CoudertMadre1991}.
Classical BDD techniques such as cofactoring and restriction operators make it possible to eliminate irrelevant parts of the input space~\cite{CoudertMadre1992}.
These operations have been extensively studied as a means to reduce decision diagram size and complexity
and could enable the storage of additional quantitative information, such as action costs.

Overall, this work shows that BDDs offer a compelling alternative to neural-network-based compression for certified avionics decision systems,
combining exactness, embeddability, execution predictability, and formal verifiability within a single unified framework.

%
%
\bibliographystyle{splncs04}
\bibliography{ref}

\end{document}